\documentclass[iop]{emulateapj}

\usepackage[colorlinks=true,linkcolor=blue,citecolor=blue]{hyperref}
\usepackage{graphicx}
\usepackage{float}
\usepackage[caption=false]{subfig}

\usepackage{natbib}

\newcommand{\DHSres}{299 - 334}
\newcommand{\DHSresApprox}{300}
\newcommand{\DHSgap}{0.04}

\newcommand{\SOSSrange}{0.6 - 2.5~$\mu$m}
\newcommand{\SOSSrangeto}{0.6 to 2.5~$\mu$m}

\shorttitle{NIRCam's DHS Mode}
\shortauthors{Schlawin et al.}

\begin{document}


\title{Two NIRCam channels are Better than One: How JWST Can Do More Science with NIRCam's Short-Wavelength Dispersed Hartmann Sensor}


\author{E. Schlawin, M. Rieke, J. Leisenring, L. M. Walker, J. Fraine, D. Kelly, K. Misselt}
\affil{Steward Observatory, Tucson AZ 85721}
\email{eas342@email.arizona.edu}

\author{T. Greene}
\affil{NASA Ames Research Center, Moffett Field, CA 94035}

\author{M. Line}
\affil{School of Earth and Space Exploration, Arizona State University, Tempe, AZ 85281}

\author{N. Lewis, J. Stansberry}
\affil{Space Telescope Science Institute, 3700 San Martin Drive, Baltimore, MD 21218, USA}



\begin{abstract}
The James Webb Space Telescope (JWST) offers unprecedented sensitivity, stability, and wavelength coverage for transiting exoplanet studies, opening up new avenues for measuring atmospheric abundances, structure, and temperature profiles. Taking full advantage of JWST spectroscopy of planets from 0.6$\mu$m to 28$\mu$m, however, will require many observations with a combination of the NIRISS, NIRCam, NIRSpec, and MIRI instruments. In this white paper, we discuss a new NIRCam mode (not yet approved or implemented) that can reduce the number of necessary observations to cover the 1.0$\mu$m to 5.0$\mu$m wavelength range. Even though NIRCam was designed primarily as an imager, it also includes several grisms for phasing and aligning JWST's 18 hexagonal mirror segments. NIRCam's long-wavelength channel includes grisms that cover 2.4$\mu$m to 5.0$\mu$m with a resolving power of $R = 1200 - 1550$ using two separate configurations. The long-wavelength grisms have already been approved for science operations, including wide field and single object (time series) slitless spectroscopy. We propose a new mode that will simultaneously measure spectra for science targets in the 1.0$\mu$m to 2.0$\mu$m range using NIRCam's short-wavelength channel.
This mode, if approved, would take advantage of NIRCam's Dispersed Hartmann Sensor (DHS), which produces 10 spatially separated spectra per source at $R \sim \DHSresApprox$.
We discuss the added benefit of the DHS in constraining abundances in exoplanet atmospheres as well as its ability to observe the brightest systems.
The DHS essentially comes for free (at no time cost) with any NIRCam long-wavelength grism observation, but the detector integration parameters have to be selected to ensure that the long-wavelength grism observations do not saturate and that JWST data volume downlink constraints are not violated. 
Combining both of NIRCam's channels will maximize the science potential of JWST, which is a limited life observatory.
\end{abstract}





\section{Introduction}

The James Webb Space Telescope \citep[JWST; e.g.][]{gardner2006SSRv} has the potential to profoundly expand and deepen the science of transiting exoplanets \citep[e.g.][]{greene2016jwst_trans}.
The wavelength coverage from 0.6$\mu$m to 28$\mu$m adds significantly to the wavelengths covered by the Hubble Space Telescope (HST) (currently 0.09$\mu$m to 1.7$\mu$m).
Its unprecedented infrared sensitivity and stable orbit at the second Lagrange (L2) point make it an excellent platform for time series observations to characterize the atmospheres of planets that transit their host stars \citep{beichman2014pasp,stevenson2016ers}.
Ground-based observatories and the Hubble Space Telescope have made important measurements of Na, K, H$_2$O and CO in a variety of hot Jupiter and hot Neptune planets \cite[e.g.][]{charbNa,redfield2008sodium,snellen2008Na209,brogi2012tauB,rodler2012taub,birkby2013water,kreidberg2014wasp43,fraine2014hatp11,deming13,sing2016continuum}.
JWST has the capability of characterizing carbon-bearing and oxygen-bearing molecules such as CH$_4$, CO, CO$_2$ and H$_2$O at high precision to improve measurements of planet metallicity and C/O in comparison to its host star.
Depending on the composition and size of the clouds and hazes in exoplanet atmospheres, the longer wavelengths observed by JWST may penetrate these clouds and hazes \citep[e.g.][]{morley2015superEclouds}.

No single JWST instrument can measure a spectrum over the full wavelength range from 0.6$\mu$m to 28$\mu$m, so multiple transits or secondary eclipses are necessary to piece together a broadband spectrum.
The largest bandwidths possible are NIRSpec's low resolution double prism ($0.7-5.0~\mu$m) and MIRI's low resolution prism (LRS; $5-12~\mu$m).
Unfortunately, NIRSpec's low resolution prism mode is so sensitive that it saturates for most sources at a $J$ magnitude of $\sim$11 \citep{beichman2014pasp}, so many nearby bright planetary systems must be observed with higher resolution (but smaller wavelength coverage) modes and instruments.
For example, to create a transmission spectrum for HD~209458b \citep{henry00,charbonneau00} from 0.7$\mu$m to 12$\mu$m, an observer would have to stitch together at least four different transits:\ one with NIRISS SOSS (\SOSSrange), two with NIRCam's long-wavelength channel grisms (F322W2 and F444W + grisms; $2.44-4.98~\mu$m)\footnote{Or, alternatively NIRSpec's G235H and G395H modes ($1.7-5.0~\mu$m.} and one with MIRI LRS ($5-12~\mu$m).
Using the Astronomer's Proposal Tool (APT) software\footnote{http://www.stsci.edu/hst/proposing/apt}, each transit observation (pre-ingress, transit, and post-ingress) requires $\sim$9~hours and incurs overheads of $\sim$2~hours, resulting in a total of $\sim$11~hours per transit and hence $\sim$44~hours total to cover
the full wavelength range.

The NIRCam instrument is primarily an imager, but it includes several spectroscopic modes originally designed for observatory wavefront sensing.
These include the grisms installed in its long-wavelength channel as well as the Dispersed Hartmann Sensors (DHSs) installed in its short-wavelength channel.
The long-wavelength grisms are used in concert with broadband filters.
The F322W2 filter covers 2.44$\mu$m to 4.02$\mu$m half maximum throughput while the F444W filter covers 3.88$\mu$m to 4.98$\mu$m at half maximum throughput.
These filters and grism are enabled and supported JWST science modes.
The DHS would be used primarily with the F150W2 filter, which covers from 1.01$\mu$m to 2.35$\mu$m at half maximum throughput.
The range from 2.02$\mu$m to 2.35$\mu$m will probably be discarded because the second order of the spectrum begins to overlap at 2.0$\mu$m.\footnote{The prisms that spatially separate the 10 DHS spectra cross-disperse the second order spectrum by 0 to 15 pixels, so the contamination can be reduced in some cases.}
When using the DHS with the F150W2 filter and pairing it with the long wavelength grisms (F322W2 and F444W), the wavelength coverage will therefore be 1.0 to 5.0$\mu$m with a gap from 2.00$\mu$m to 2.44$\mu$m.
This wavelength gap can include a CH$_4$ feature for cool planets, but other CH$_4$ bands are also visible at 1.7$\mu$m and 3.5$\mu$m.


The NIRCam instrument can reduce the number of transits needed to build up a 1$\mu$m to 5$\mu$m transmission or emission spectrum of a planet.
This is because the instrument has a dichroic that splits the incoming beam into a short-wavelength channel and long-wavelength channel, so simultaneous imaging and/or spectroscopy are possible at short ($<$2.4$\mu$m) and long ($>$2.4$\mu$m) wavelengths.
In principle, the DHS can be used to simultaneously cover 1.0$\mu$m to 2.0$\mu$m while the long-wavelength grisms cover either 2.44$\mu$m to 4.02$\mu$m (with the F322W2 filter) or 3.88$\mu$m to 4.98$\mu$m (with the F444W) filter.
Thus, the DHS data come ``for free'' with any long-wavelength NIRCam grism data, aside from readout and data rate considerations discussed in Section \ref{sec:readout}.
For the example above of HD~209458b, simultaneous DHS data would reduce the required transits from four to three, saving about 11 JWST hours 
if the wavelengths from 2.0$\mu$m to 2.4$\mu$m and higher Signal to Noise ratios (SNRs) from NIRISS are not required.
The DHS is expected to collect similar photon levels as HST's WFC3 grisms because the DHS pupil coverage ($\sim$30\%) and relative throughput ($\sim$30\% versus $\sim$40\% with WFC3 \texttt{http://etc.stsci.edu}) are together similar to the ratio between telescope collecting areas.
Therefore, the DHS is like a free coordinated HST observation for every long-wavelength NIRCam grism transit but without the interruptions or thermal consequences of HST's low earth orbit.

Since the DHS was not originally designed for science observations, it is not yet an approved mode for the James Webb Space Telescope science operations.
In this white paper, we explain the value of the DHS mode to fully take advantage of two NIRCam channels to maximize JWST's science potential for transiting planets as well as discuss the specifics of its implementation.
In section \ref{sec:implementation}, we discuss how the DHS would work for time-series observations and the implications of reading out several detectors simultaneously at a high frame rate.
In section \ref{sec:addedScience} we discuss the added science that can be achieved with the DHS, including the brightness limits and improved precision in retrieved parameters for a representative system.

\section{DHS Implementation}\label{sec:implementation}

\subsection{DHS properties}

\begin{figure}
\centering
\includegraphics[width=1.0\columnwidth]{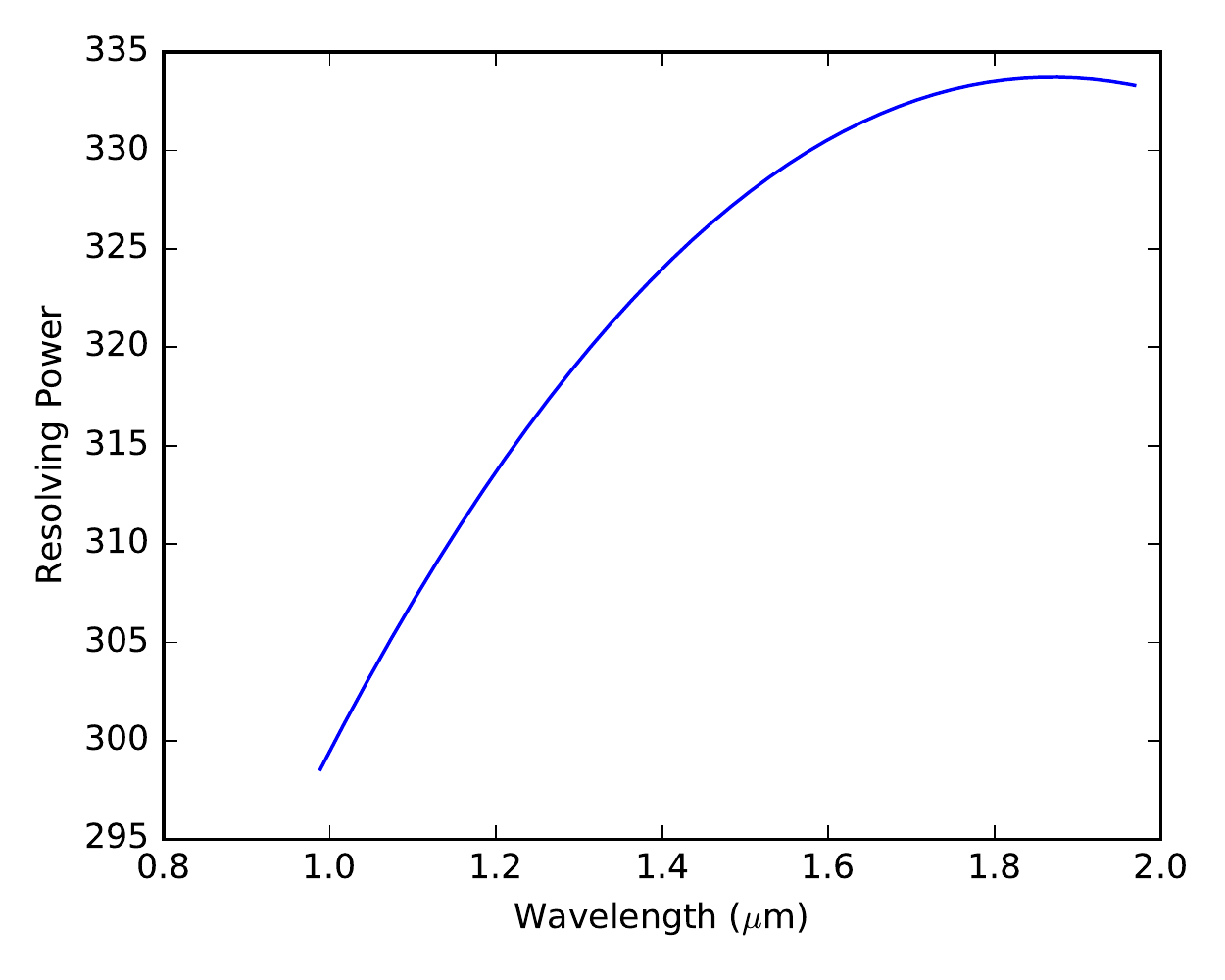}
\caption{Resolving power as a function of wavelength for the DHS.}\label{fig:DHSRes}
\end{figure}

\begin{figure}
\centering
\includegraphics[width=1.0\columnwidth]{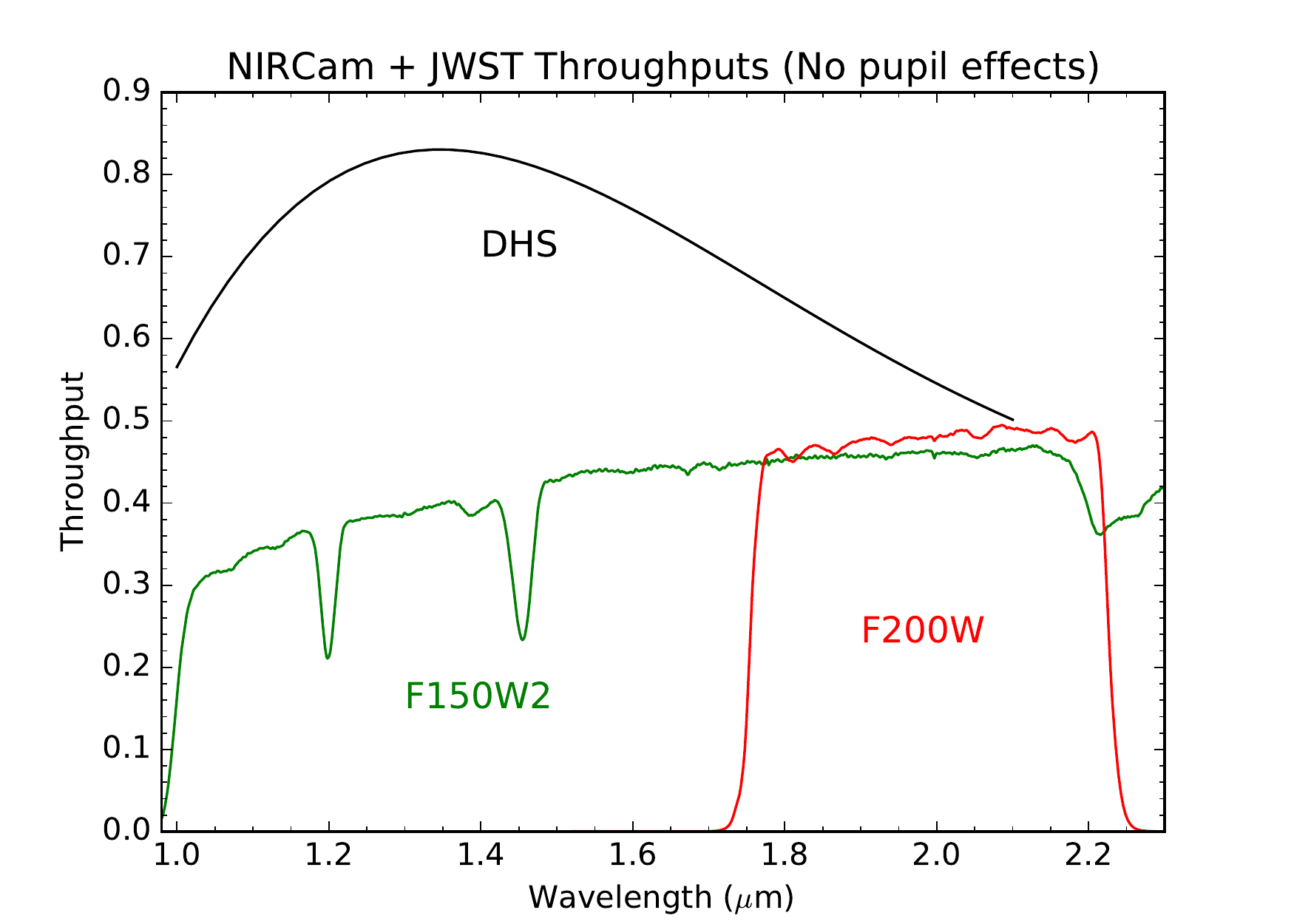}
\caption{Throughput as a function of wavelength for the DHS grism, F150W2 filter and F200W filter that includes the telescope and instrument transmission. The total throughput is the product of the filter curve, ``DHS'' grism and the pupil area fraction from Table~\ref{tab:pupfrac}..}\label{fig:DHSthrough}
\end{figure}

The DHS expands the wavelength range that NIRCam can achieve on a single observation by simultaneously obtaining spectroscopy with NIRCam's short wavelength channel and long wavelength channel.
The maximum wavelength range for the F150W2 filter in the short wavelength channel is fixed at 1.01 to 2.02$\mu$m with the region from 2.02-2.35$\mu$m contaminated by second order overlap.
The DHS are slitless, so the resolution is set by the pupil sizes of the individual DHS elements, which are approximately the same width as each other in the dispersion direction.
Figure~\ref{fig:DHSRes} shows the resolving power as a function of wavelength, estimated from a narrowly peaked input spectrum and the fits to the spatial width along the spectrum.
There is a spatial gap between short wavelength detectors which corresponds to about \DHSgap$\mu$m in wavelength.
The center of this gap depends on source position.
The DHS can be used in tandem with either the F322W2 or F444W filter for the long wavelength grism observations.
The properties of the DHS as compared to the grisms are listed in Table~\ref{tab:DHSgprop}.
If the region from 2-2.23$\mu$m is highly desirable such as for measuring the CH$_4$ feature shown in Figure~\ref{fig:HATp12spec}, the DHS can be used with the F200W filter (1.75 to 2.23$\mu$m half maximum) to partially fill in the gap between the DHS+F150W2 spectrum and the long wavelength channel (grism+F322W2 spectrum), though the DHS grism transmission is lower above 2.0$\mu$m, shown in Figure~\ref{fig:DHSthrough}. 

\begin{table*}
\centering
\caption{Simultaneous DHS and Grism Properties (Module A)}\label{tab:DHSgprop}
\begin{tabular}{lcc}
\hline \hline
 & DHS & Grism \\
\hline \hline		
Resolving Power: &  \DHSres & 1200 - 1550 \\
Dispersion: & 0.29 nm / px & 1.0 nm / px \\
Wavelengths: & 1.05 - 2.02$\mu$m (F150W2)\footnote{For a field point matched to the F322W2 filter and long wavelength grism.} & 2.44 - 4.02$\mu$m (F322W2)\\
			 & 1.75 - 2.23$\mu$m (F200W)\footnote{For a field point matched to the F322W2 filter and long wavelength grism.} & 2.44 - 4.02$\mu$m (F322W2)\\
			& 	1.01 - 1.94$\mu$m (F150W2) \footnote{For a field point matched to the F444W filter and long wavelength grism. There is a \DHSgap$\mu$m gap between detectors that depends on field position.}	& 3.88 - 4.98$\mu$m (F444W) \\
Plate Scale: & 32 mas / px  &  65 mas / px\\
Undeviated Wavelength: & 1.36$\mu$m & 3.94$\mu$m   \\
\hline
\end{tabular}
\end{table*}

\subsection{DHS Layout}\label{sec:layout}

\begin{figure*}
\centering
\includegraphics[width=0.85\textwidth]{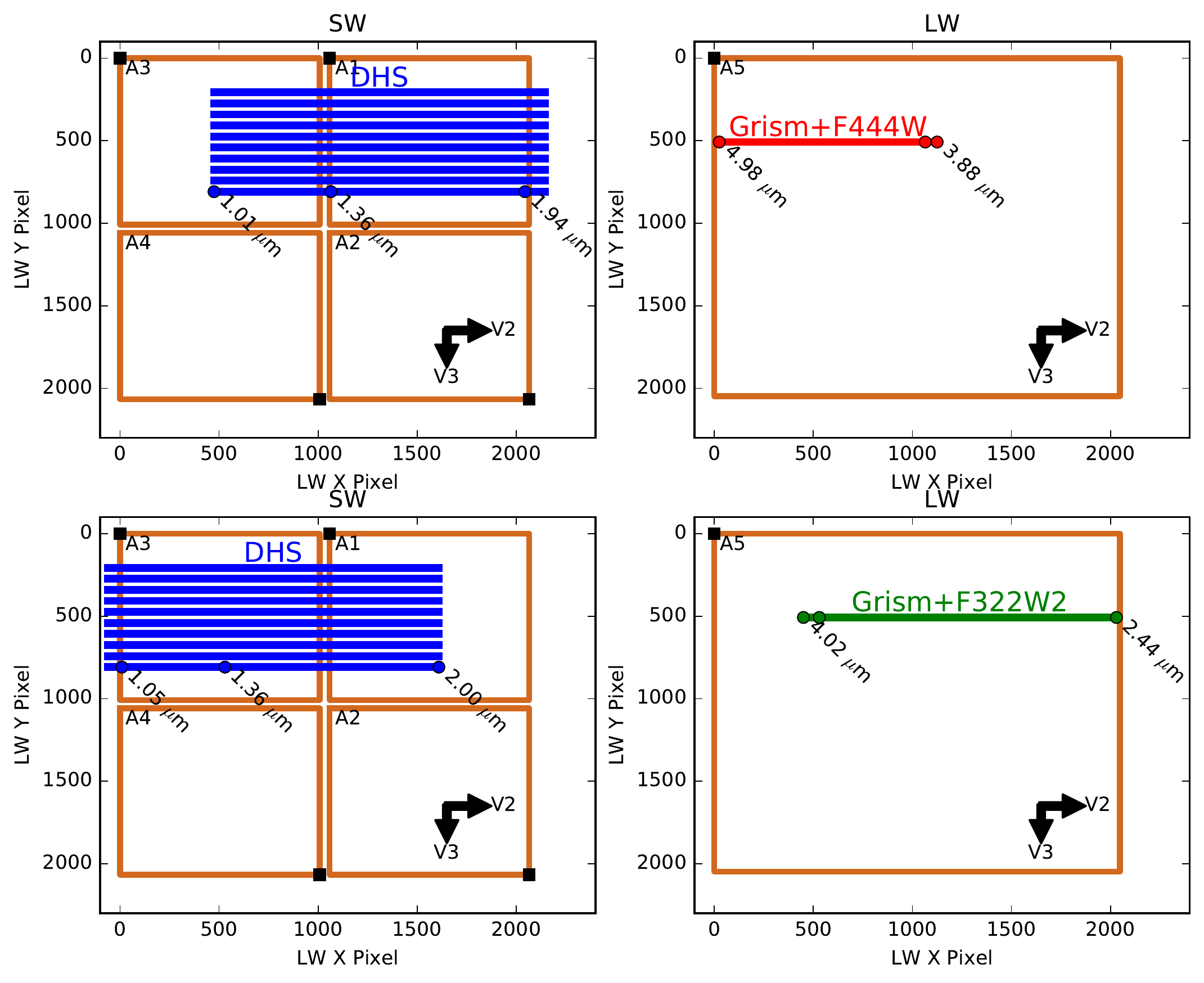}
\caption{DHS wavelength coverage (when using the F150W2 filter) for two field positions that will permit grism spectroscopy simultaneously with the F322W2 and F444W filters in the long wavelength channel for the same input source. The V2/V3 axes show the JWST observatory axes and the detectors (A1 - A5) are shown with their (0,0) corners as black boxes. The circles plotted for each grism are the shortest usable wavelength, undispersed wavelength and longest usable wavelength.}\label{fig:DHSwaveWGrism}
\end{figure*}

A point source observed with the DHS and grism modes of NIRCam will produce 10 spectra on the short-wavelength channel and one spectrum on the long-wavelength channel -- see Figure~\ref{fig:DHSwaveWGrism}.
The DHS spectra span two of NIRCam's short-wavelength detectors (covering 1.02$\mu$m to 2.02$\mu$m when using the F150W2 filter) with a \DHSgap$\mu$m gap due to the separation between detectors.
The DHS mode is well suited for one bright source because it spreads the light over the 10 separate spectra, but is ill-suited for a crowded field, where the spectra would overlap.
Each of the DHS spectra are separated in the spatial direction by an average of 123 pixels, so a sub-array of at least 1300 pixels must be extracted to gather all DHS flux for a single source.

\begin{figure}[!t]
\centering
\includegraphics[width=1.0\columnwidth]{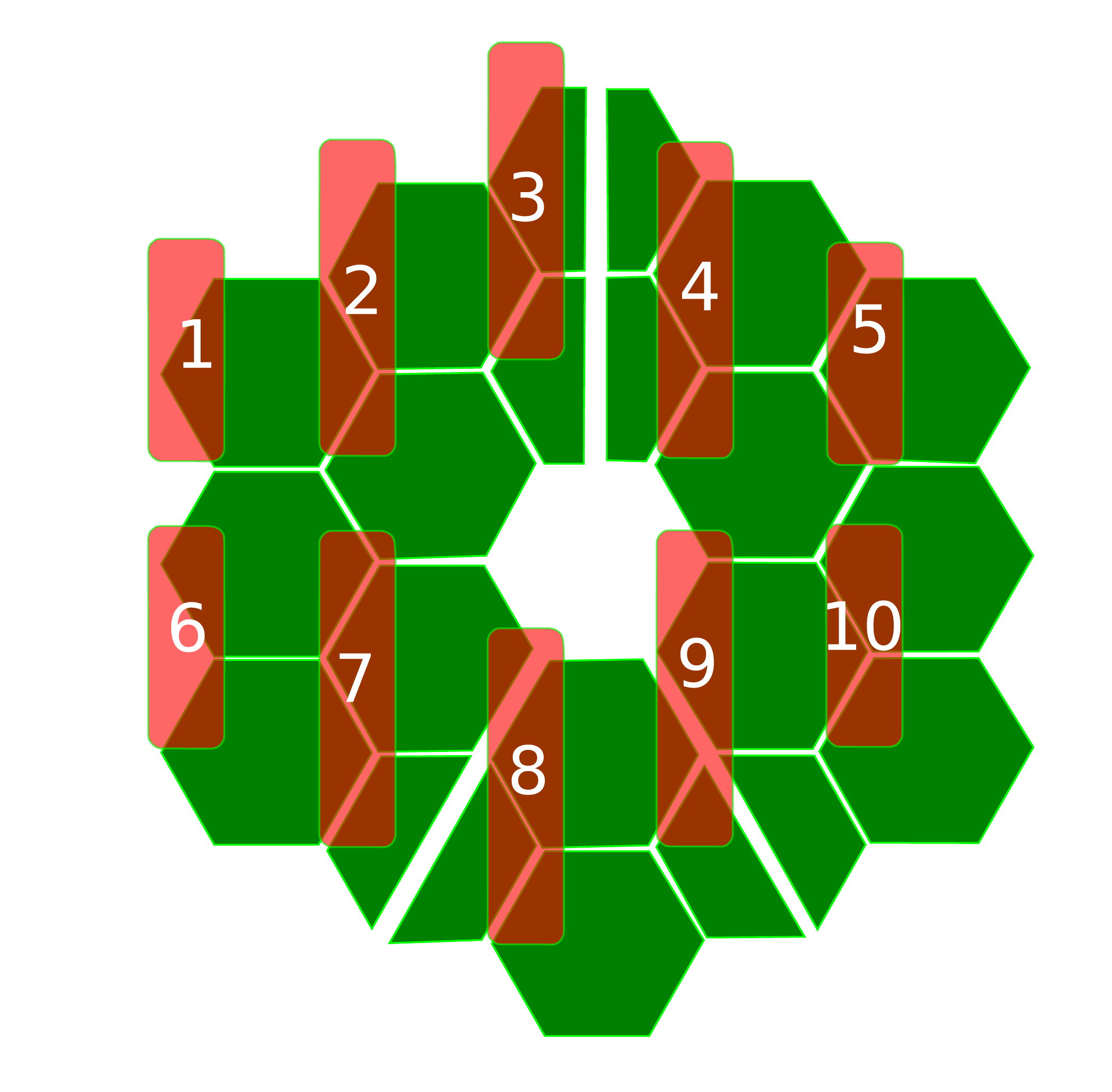}
\caption{DHS pupil coverage by each of the 10 DHS grisms, with the fractional throughput given in Table~\ref{tab:pupfrac}. 
The throughput is modulated by the size of the DHS grism as well as whether it coincides with a secondary mirror support (inverted ``Y'' shaped feature).
The pupil wheel is clocked to make the DHS spectra straight along rows.
This moves the DHS elements off-center from JWST's pupil and thus decreases the throughput for elements 1 and 6 substantially, but the overall combined throughput for all 10 elements is only reduced from 32\% to 29\%.}\label{fig:DHSvsPupilOverlay}
\end{figure}

Each DHS spectrum covers roughly 4\% of the JWST pupil, though they vary in size and throughput depending on where they lie relative to the secondary mirror strut supports, seen in Figure~\ref{fig:DHSvsPupilOverlay}.
We elect to clock the pupil wheel by 1.9$^\circ$ in order to align the spectra with rows on the short wavelength detector.
Aligning spectra along the rows enables small stripe mode readouts (such as 2048 $\times$ 64) to read one DHS spectra simultaneously with long wavelength grism spectra.
This is needed for bright targets that will saturate the long wavelength detector with longer frame times, as discussed in Section \ref{sec:readout}.
There is a price for clocking the DHS pupil wheel to align spectra with rows: all DHS grisms are partially obscured by either the secondary supports or the outer edges of the primary, save elements 7 and 10.
Fortunately, the relative obscuration for all 10 DHS spectra is small, going from 32\% total throughput when the pupil wheel is centered (which produces tilted spectra), versus 29\% if the pupil wheel is clocked to align the spectra along rows.

\begin{table}[!b]
\centering
\caption{DHS Pupil Coverage}\label{tab:pupfrac}
\begin{tabular}{cc}
\hline \hline
Aperture \# & Pupil Fraction \\
\hline \hline
1  & 0.012 \\
2  & 0.032 \\
3  & 0.032 \\
4  & 0.040 \\
5  & 0.021 \\
6  & 0.015 \\
7  & 0.042 \\
8  & 0.040 \\
9  & 0.036 \\
10 & 0.022 \\
\hline
Total & 0.292\\
\hline
\end{tabular}
\tablenotetext{0}{The pupil coverage of the 10 DHS grisms compared to the total primary collecting area for the pupil wheel position that aligns DHS spectra with detector rows.}
\end{table}

The throughput for DHS spectroscopy is the product of the DHS grating efficiency, telescope and instrument throughput for a given filter (both shown in Figure~\ref{fig:DHSthrough}) and the pupil coverage by the DHS elements listed in Table~\ref{tab:pupfrac}.
A fractional pupil of 1.00 would use the entire pupil and thus take advantage of the maximum JWST collecting area.
For a sub-array size encompassing only 1 DHS spectrum (such as a 2048 $\times$ 64 subarray), the largest unobscured DHS element should be used - number 7 in Figure~\ref{fig:DHSvsPupilOverlay}.
Only the long-wavelength grisms of Module A are considered here, because those in Module B lack an anti-reflective coating.



\subsection{Readout Considerations}\label{sec:readout}
\begin{figure}[t]
\centering
\includegraphics[width=1.0\columnwidth]{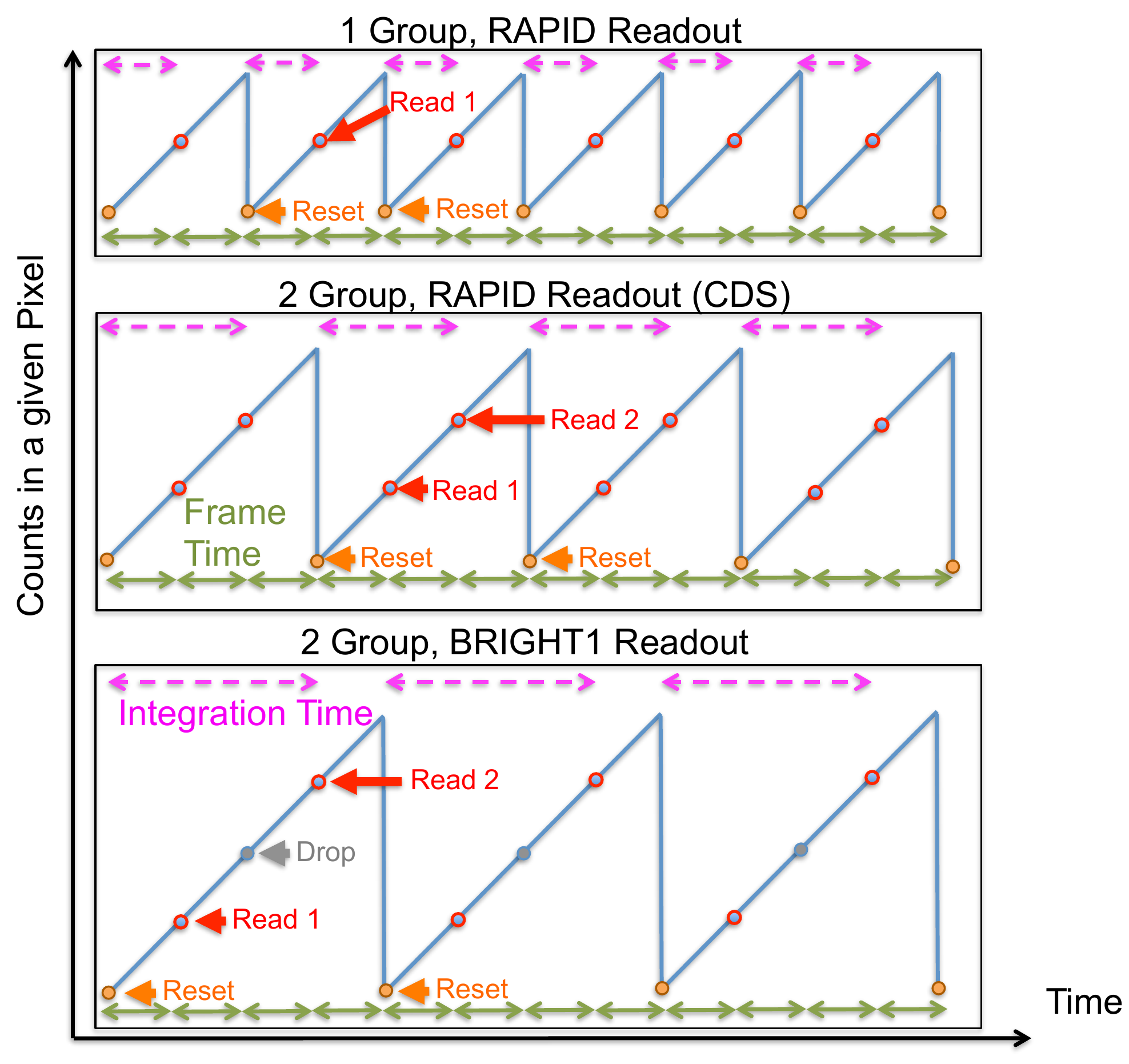}
\caption{Schematic of example readout modes for an exposure that is 12 frame times in duration.
A single exposure consists of 1 or more groups per integration.
No reset frames or dropped frames are saved. 1~Group, RAPID readout can be used to increase the saturation limits provided that drifts between reset levels are sufficiently stable.
The BRIGHT1 readout increases the observing efficiency and reduces data volume over CDS but has a longer minimum integration time, resulting in fainter saturation limits.}\label{fig:readout}
\end{figure}

NIRCam's detectors have numerous pixel readout schemes that factor into saturation limits and SNR calculations for a given observation.
We adopt the following nomenclature for the readout (Figure~\ref{fig:readout}):\ an \emph{integration} (also commonly refered to as a ``ramp'') consists of a pixel-reset frame followed by one or more groups of non-destructive reads to sample the pixels' charge accumulation. An \emph{exposure} is considered to be a collection of multiple integrations of the same type and length. Finally, an \emph{observation} consists of many exposures.

The NIRCam instrument supports 9 dedicated readout modes, each with a set number of reads up the ramp co-added on board (collectively called a group) and a set number of reads dropped between the groups, both of which reduce data volume.
As the exoplanet transit science case will be focused on bright targets, we consider two of the allowed modes in this paper: RAPID (all read frames up the ramp are saved and downlinked) and BRIGHT1
(alternating reads up the ramp dropped, i.e. reset-read-drop-read-drop-...). 
In these cases, the number of coadded frames is 1, so the number of groups is the same as the number of reads.
A general discussion of HAWAII-2RG detector readouts can be found in  \citet{rauscher2007detectors} and
a detailed description of the 9 readout patterns implemented for NIRCam can be found
in  \citet{roberto2009optimalR1,roberto2010optimalR2}.

We consider a simplified example of a time series of 4 integrations using 2 Groups in RAPID mode (a standard correlated double sampling, CDS pattern) illustrated in Figure~\ref{fig:readout} and assume an approximate frame time of 10.0 seconds, close to the actual 10.74 seconds full frame read time.
CDS is equivalent to RAPID mode with 2 reads up the ramp.
The raw data delivered to observers for use in a pipeline would contain a data cube.
If the data is in full frame mode, the cube would have dimensions of 2048 $\times$ 2048 $\times$ 8, which can be thought of as 8 image planes.
In this example there are 4 integrations, each consisting of a reset followed by two read frames, which are saved as two image planes per integration.	
For simplicity in this example, we ignore many of the corrections that are applied in a pipeline such as non-linearity correction, reference pixel correction, intrapixel capacitance, flat fielding etc.
Under these assumptions, the flux (in DN/s) in any integration is the difference between the two reads divided by the frame time.
This reduces the 8 image planes, to 4 flux planes, one per integration.
Spectral extraction on the flux planes \citep[e.g.][]{horne1986optimalE} then produces the flux per wavelength bin.
A final reduced time series would consist of flux values at each given wavelength bin for each of the 4 time coordinates centered on 15 sec, 45 sec, 75 sec and 105 sec relative to the beginning of the observation sequence.

The BRIGHT1 and RAPID modes give high time resolution and allow one to detect and flag cosmic ray hits without having to discard large amounts of data.
The number of groups per integration is a variable set in a proposal with APT.
When the number of groups is 1 (``1~Group Readout''), there is no ability to subtract out the current bias frame so it relies on the bias level being constant through the whole time series.
When the number of groups is 2, the slope can be calculated from the difference between the two reads divided by the group time.
It is the same as the CDS mode described above when using RAPID readout.
For a higher number groups, a line is fit to the the ramp to calculate a slope image for every integration.
In exoplanet time series, the high signal to noise ratios required generally mean that bright sources will be observed so these will generally not benefit from skipped frames or co-added frames of other JWST readout modes, which come at the expense of lower time resolution, reduced ability for cosmic ray rejection and fainter saturation limits.

\begin{table}[!b]
\centering
\caption{Grism \& DHS Performance}\label{tab:SatSNRsubA}
\begin{tabular}{llrr}
\hline \hline
Sub-Array Size &  LW Saturation & \# of DHS & DHS Noise\\
               &  K Magnitude   &           & ppm \\
\hline \hline
2048$\times$ 64 & 4.52 & 1 &  72 -- 110 \\
2048$\times$ 128 & 5.26 & 1 &  100 -- 160 \\
2048$\times$ 256 & 6.01 & 2 &  100 -- 160 \\
2048$\times$ 2048 & 8.26 & 10 & 160 -- 270 \\
\hline
\end{tabular}
\tablenotetext{0}{G2V-type star saturation K band limits are shown for the current available array sizes in APT, which determine the number of spectra from DHS elements that can be included.
The noise in the DHS spectra over 1.1 -- 1.9$\mu$m at the given saturation limit for a 1-hour transit is shown in column 4 at native R $\sim$ \DHSresApprox\ resolving power.}
\end{table}

\begin{figure}[!t]
\includegraphics[width=1.0\columnwidth]{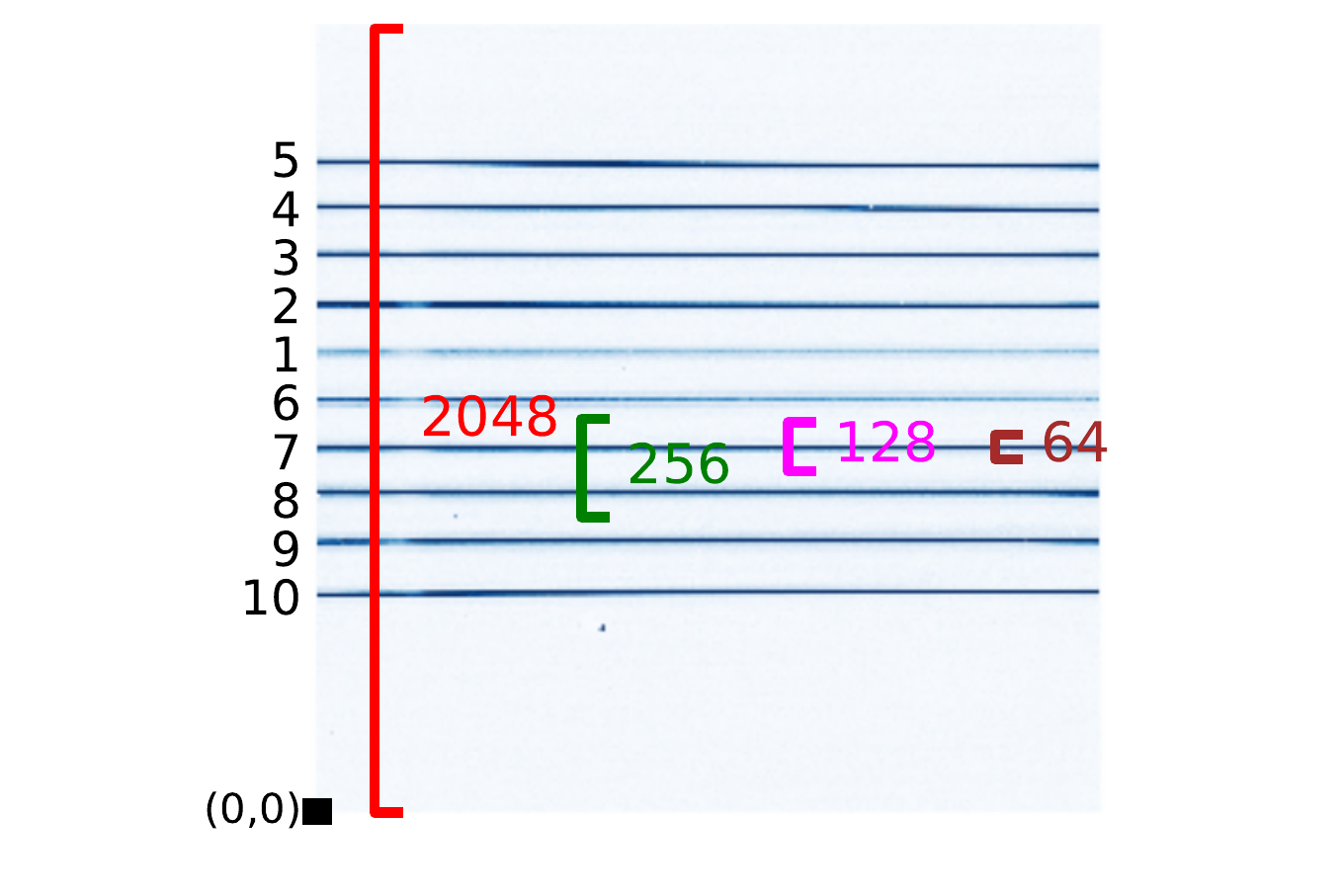}
\caption{Example DHS image on the A1 short wavelength detector taken during cryogenic vacuum testing of the flight hardware -- see Figure~\ref{fig:DHSwaveWGrism} for the detector's position.
Each element is labeled corresponding to the DHS apertures shown in Figure~\ref{fig:DHSvsPupilOverlay} and pupil fraction in Table~\ref{tab:pupfrac}.
The sub-array size options currently available (all 2048 pixels wide) are shown with selections for the brightest spectra available - elements 7 and 8.}\label{fig:DHSaps}
\end{figure}

Each NIRCam array must be read out with identical window sizes and exposure times due to restrictions in the Onboard Scripting Software (OSS).
This means that some balance must be made between saturating the grism spectrum (which has much higher throughput) in the long-wavelength and getting the most flux in the short-wavelength DHS spectra (which have lower throughput and also spread the light over more pixels).
The current time series observing templates available in APT are limited to subarray sizes of 2048 $\times$ 64, 2048 $\times$ 128, 2048 $\times$ 256 and 2048 $\times$ 2048 (full frame), which would respectively cover 1, 1, 2 and 10 DHS spectra as shown in Figure~\ref{fig:DHSaps}.
These are the subarrays that will be available to Cycle 1 science programs and likely stay this way for continuity through the mission in order to stick to detector ``sweet spots'' that are well characterized and calibrated.
In principle, the short wavelength detectors can be read out with different sizes and read patterns from long wavelength array or even what is available in APT, but the science benefits must be traded off with operational costs to implementing and testing new modes in the OSS software and data handling.

We recommend that the sub-array locations be adjusted such that the short-wavelength sub-array can be centered on one DHS spectrum or a subset of the 10 and the long-wavelength array can be centered on a single grism spectrum. For the 2048~$\times$~64 and 2048~$\times$~128 subarrays, which can only encompass one spectrum with more than 10 pixels of background, they should be centered on the spectrum from element 7, since it has the maximum possible throughput listed in Table~\ref{tab:pupfrac}. For the 2048~$\times$~256 subarray, the location should be centered between the spectra from elements 7 and 8 since they have the highest throughput of any adjacent pairs from Table~\ref{tab:pupfrac}. These recommended subarray positions are illustrated in Figure \ref{fig:DHSaps}.

The balance between saturating the long-wavelength grism and getting enough SNR in the DHS is listed in Table~\ref{tab:SatSNRsubA}.
The table shows the brightest observable G2V star for a given array size in the long-wavelength grism and the DHS noise at 1.1 to 1.9 $\mu$m for a G2V type star at this brightness observing a 1 hour long transit and 1 hour on a star out of transit, binned to the native resolving power of R $\sim$ \DHSresApprox.
We use the standard correlated double sampling (CDS) readout mode shown in Figure~\ref{fig:readout} and we consider 80\% well depth in the second read as saturation.
If the bias level after each reset remains stable over the course of observations, it may be possible to use ``1~Group Readout'', illustrated in Figure~\ref{fig:readout}.
The 1~Group Readout method can push the brightness limits by a factor of 2 over CDS because it reduces the integration time by a factor of two.
The sub-array sizes and order of DHS elements is shown in Figure~\ref{fig:DHSaps}.
For the brightest sources, which require the 2048 $\times$ 64 pixel subarray, the DHS throughput loss is significant, because only 1 DHS spectrum out of 10 is used, meaning only 4.2\% of JWST's 25 m$^2$ collecting area is used.

\subsubsection{Data Volume}
\begin{figure}[!t]
\includegraphics[width=1.0\columnwidth]{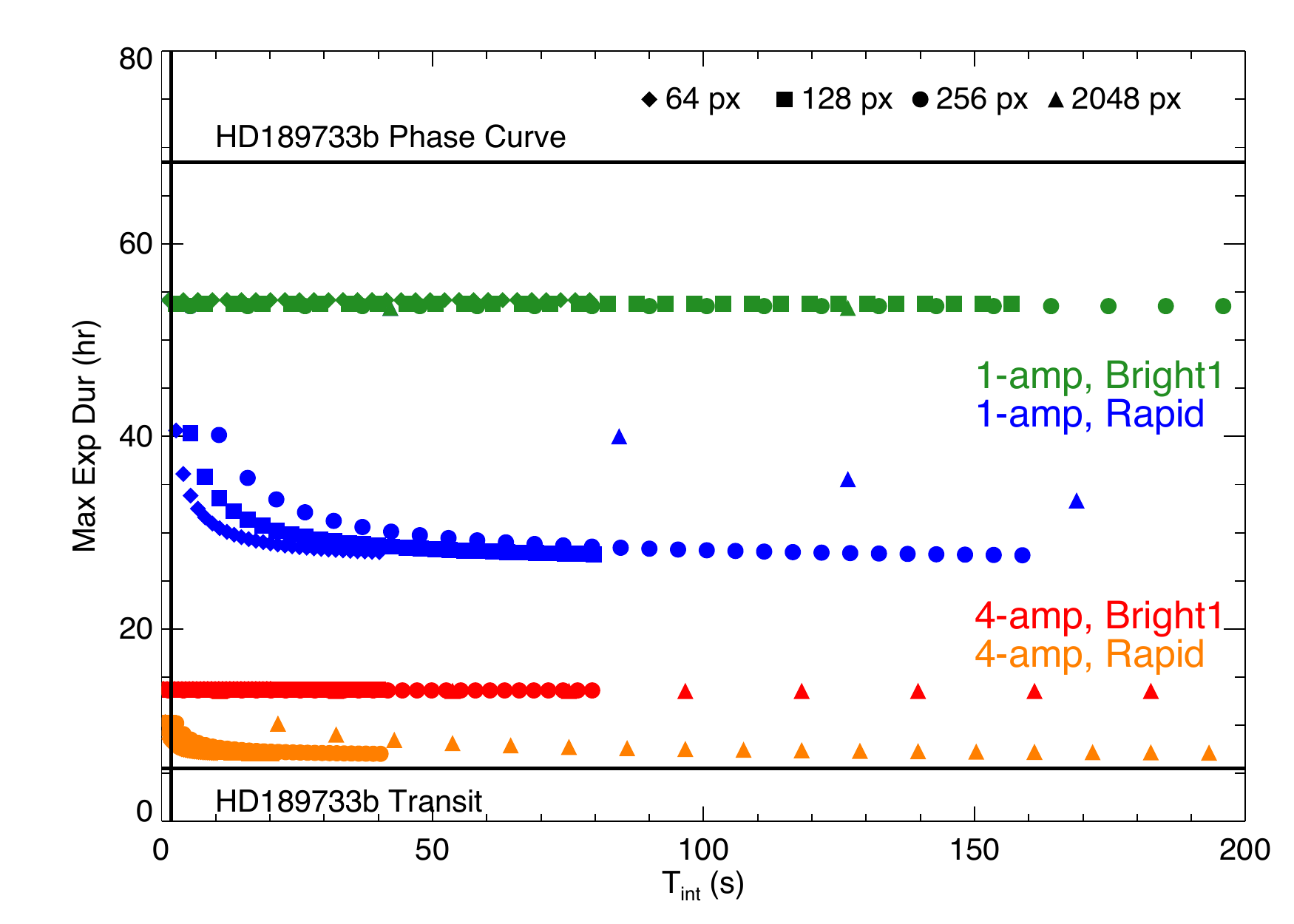}
\caption{The maximum time JWST NIRCam can expose three detectors (e.g. A1, A3 and A5 in Figure~\ref{fig:DHSwaveWGrism}) before filling up JWST's Solid State Recorder (SSR). The readout modes shown are for RAPID readout and BRIGHT1 readout for both 1 amplifier and 4 amplifiers. The number of pixels read out affects the integration time and thus the data rate. The time needed for a phase curve of the planet HD 189733b and a transit of the system are shown for reference.}\label{fig:DataVolume}
\end{figure}

For most transiting planet science, the targets are bright so the pixels will be read out with very few or zero co-added and dropped frames.
The amount of data streaming from the detectors can overload the Solid State Recorder (SSR), which has a capacity of 540 Gbits \citep{johns2008L2comm}.
We consider the use of 458 Gbits of space for science data as the maximum the SSR can handle at once to allow for storage of telemetry.
This overload can happen in as few as 10.3 hours if three NIRCam detectors are used simultaneously and all are read out with four amplifiers in CDS mode (meaning no samples are co-added or dropped) -- see Figure~\ref{fig:DataVolume}.

In normal operations, science and telemetry data from the James Webb Space Telescope are downlinked in a 4 hour contact time with an antenna in the Deep Space Network \citep{dashevsky2008groundflight}.
There will be two downlinks per day so on average the total data rate can be no more than $\sim$458 Gbits per 12 hours.
Science operations can continue during the 4 hour contact time, but re-pointing the high gain antenna can cause $<$0.07 \arcsec\ disturbances \citep{beichman2014pasp}.
To satisfy these normal operations, the data volume is a concern for bright objects requiring longer than $\sim$10 hours of continuous spectroscopy when using 3 detectors and 4 amplifiers simultaneously.

There are a few ways to reduce the data volume streaming from the detectors.
One way is to read each detector with a single amplifier.
This reduces the frame rate by a factor of four (for the four amplifiers), thus allowing $\sim$40 hours before filling JWST's solid state recorders.
The lower frame rate, however, means that the detector saturates at 1/4 the flux as four amplifiers.
Second, the detector reads can be skipped or co-added such as in BRIGHT1 mode, seen in Figure~\ref{fig:readout}.
In BRIGHT1 mode, a frame is dropped between read 2 and read 1 to increase the integration time and efficiency while reducing data volume.
This only works for sources faint enough that the long-wavelength grisms do not saturate.
Perhaps future changes to the OSS could allow operating the LW detectors with 4 outputs and the SW ones with 1 output, reducing data volume considerably.

\section{Science Benefits with the DHS}\label{sec:addedScience}

\subsection{Brightness limits}\label{sec:brightness}

\begin{figure}[!h]
\includegraphics[width=1.0\columnwidth]{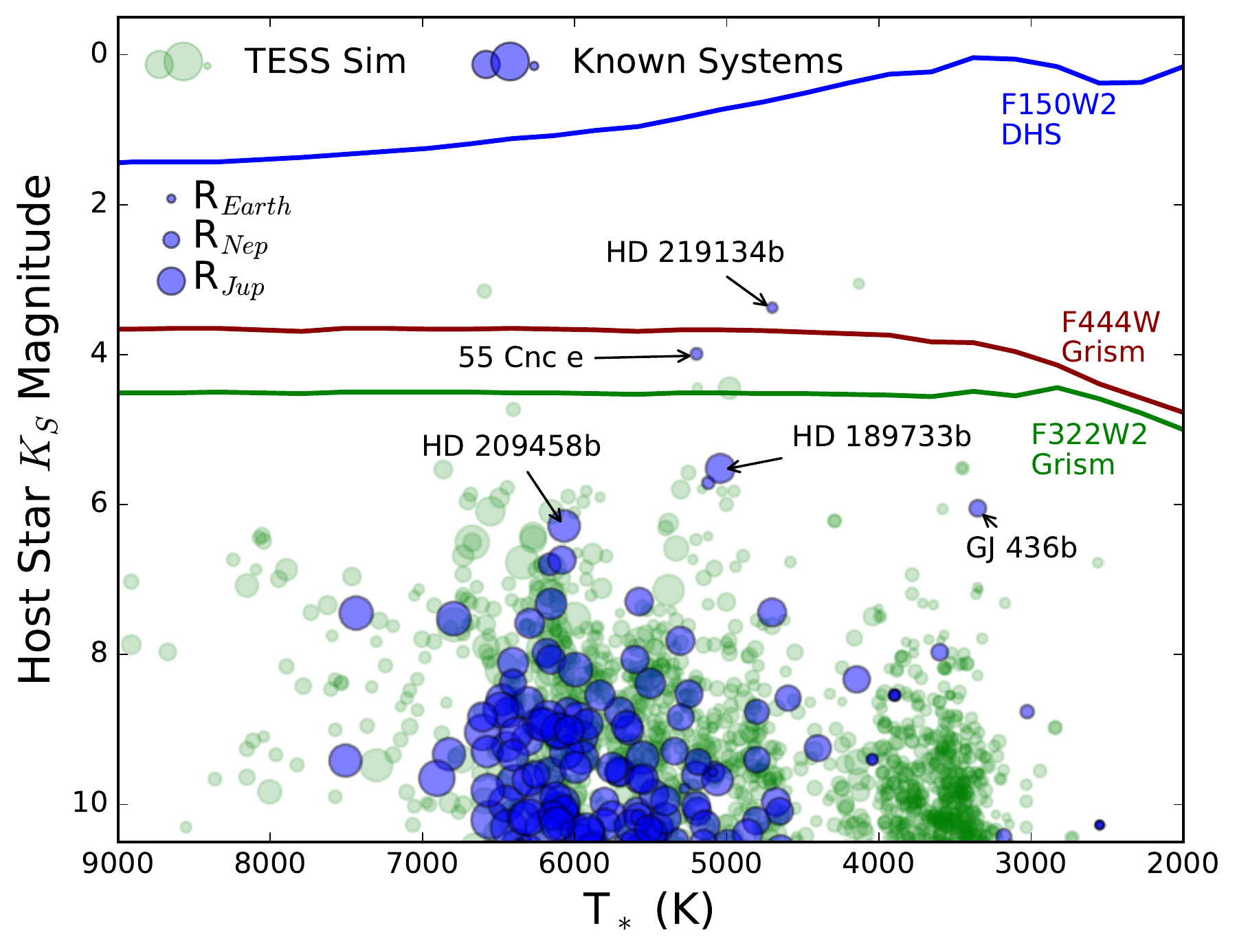}
\caption{Saturation limits of the DHS mode (blue line), the long-wavelength grism + F322W2 filter combination (green line) and the long-wavelength grism + F444W filter (dark red line) as a function of stellar effective temperature.
The magnitudes are Vega $K_S$ magnitudes where 80\% of the well depth is filled in the standard CDS readout mode (described in Figure~\ref{fig:readout}) for a 2048 $\times$ 64 pixel sub-array using 4 amplifiers.
The blue points are known transiting systems and the green points are simulated TESS planets \citep{sullivan2015tess}.}\label{fig:saturationLim}
\end{figure}

The DHS has the brightest saturation limits of any spectroscopic mode on JWST because each of the 10 DHS elements samples only 1\% to 4\% of the pupil and the light is dispersed over two short wavelength detectors.
Figure~\ref{fig:saturationLim} shows the saturation limits for the DHS mode along with NIRCam's long-wavelength grisms and the known transiting systems from \texttt{exoplanets.org} \citep{han2014exopDorg}, plus \citet{motalebi2015hd219134b}.
The saturation limits are found using \texttt{pysynphot} \citep{lim2015pysynphot} with Phoenix stellar models \citep{allard2012phoenix} ranging from 4.0 to 5.0 in log(g).
Any known transiting planet system can be observed with the DHS, including the super-Earth  55 Cnc e \citep{mcarthur2004disc55cnce} at $K$=4.0, which saturates the long-wavelength grism with the F322W2 filter and the brightest known transiting system HD 219134b \citep{motalebi2015hd219134b} at $K$=3.4.
The DHS also enables characterization of any future targets found in this brightness range, such as by the Transiting Exoplanet Survey Satellite (TESS) \citep{ricker2014tess}.
We plot the expected yield of TESS \citep{sullivan2015tess} in Figure~\ref{fig:saturationLim} and none of the simulated targets saturate the DHS.

\begin{figure*}[!thb]
\centering
\subfloat[HD 209458b]{
	\includegraphics[width=0.5\textwidth]{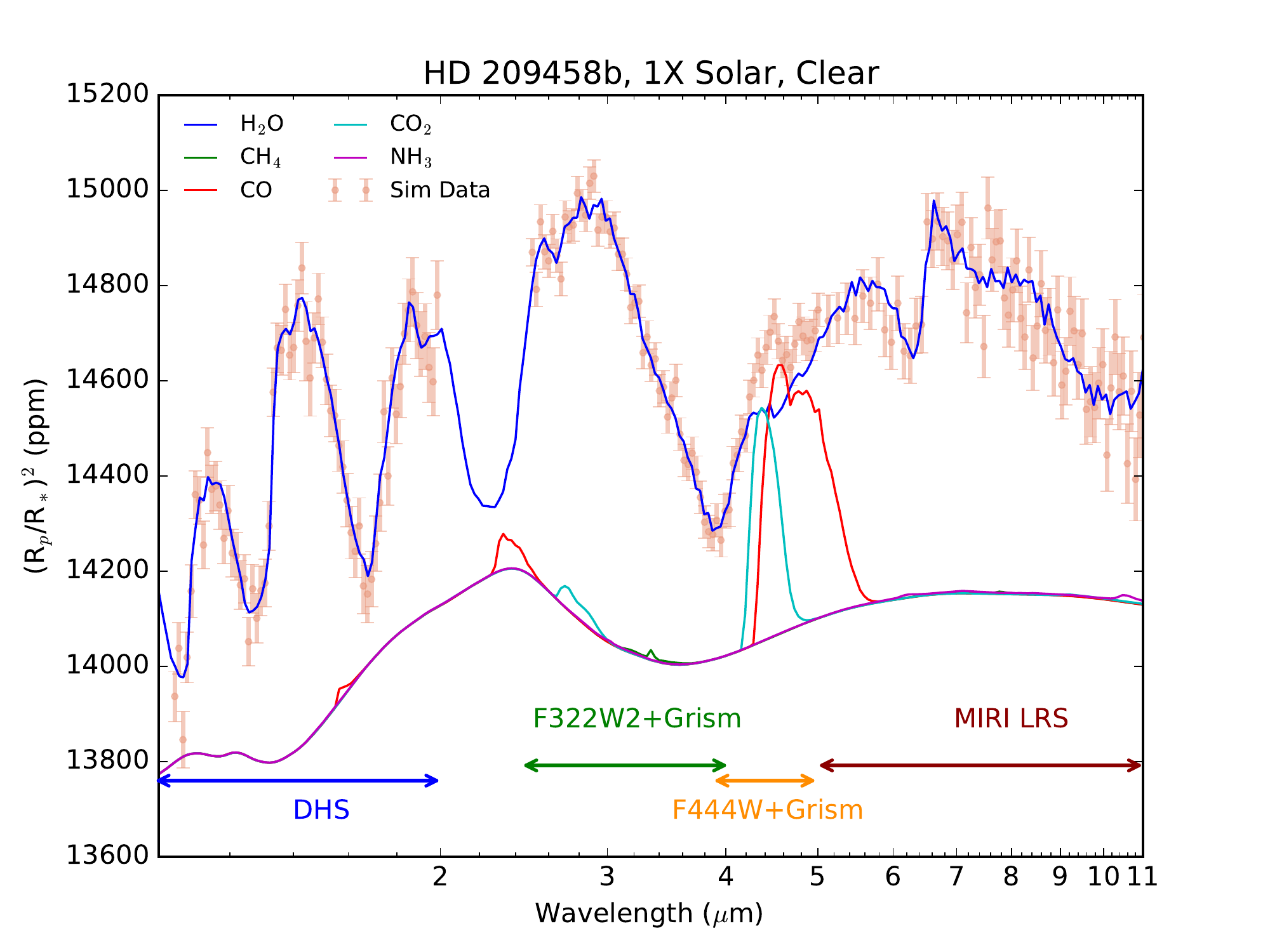}
	\label{fig:209spec}
	}
\subfloat[HAT-P-12b]{
	\includegraphics[width=0.5\textwidth]{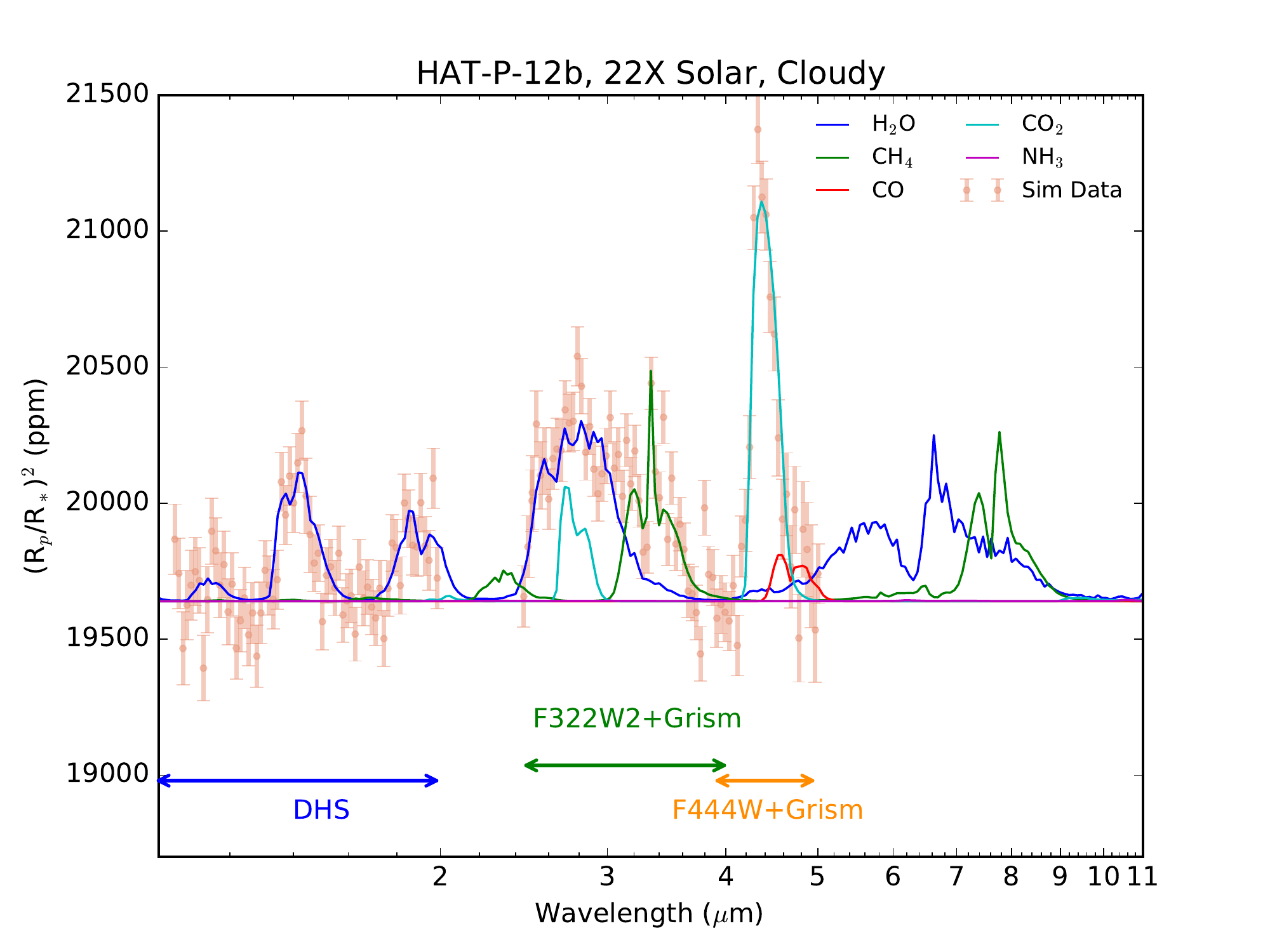}
	\label{fig:HATp12spec}
	}
	\caption{Model spectra separated by molecule for HD~209458b and the inflated warm Saturn HAT-P-12b. The new DHS mode would allow simultaneous spectra at 1-2$\mu$m ``for free'' when observing with either NIRCam's F322W2 or F444W slitless grism observations in its long-wavelength channel. a) The simulated data for HD~209458b consist of one transit in each of the long-wavelength grism modes, each accompanied by simultaneous DHS grism spectra. The observations longer than 5$\mu$m are simulated for MIRI's Low Resolution Spectrometer. b) HAT-P-12b is an inflated warm Saturn with an opaque cloud deck at 1 mbar. Error bars shown are for simulated observations of 2 total transits: one transit with the DHS and F322W2+grism together and one transit with the DHS and F444W+grism together for a total of about 17 hours of JWST time. A planet metallicity of 22~$\times$~solar was assumed from the relation in \citet{kreidberg2014wasp43}}
	\label{fig:spec}
\end{figure*} 


\subsection{CHIMERA Models}\label{sec:models}

We use the CHIMERA atmospheric retrieval suite \citep{line2013chimera,line2014CtOsecE} to illustrate the scientific value of using the DHS simultaneously with NIRCam's long-wavelength grisms.
We first simulate a forward model of a transmission spectrum for two planets, HD~209458b \citep{henry00,charbonneau00} and HAT-P-12b \citep{hartman2009hatp12}.
We then add Gaussian noise using the SNR estimated on the target to create a simulated spectrum.
Finally, we run a Markov Chain Monte Carlo (MCMC) retrieval of the atmospheric parameters for the simulated spectrum to try to recover the initial abundances and determine the uncertainties in those abundances, as in \citet{greene2016jwst_trans}.


To construct the model atmospheres and generate a spectrum we require a thermal profile and molecular abundances. For simplicity, we assume isothermal atmospheres at the planetary equilibrium temperatures. Molecular volume mixing ratios are assumed to be in thermochemical equilibrium at a represenatitive pressure (0.01 bars) and temperature using the NASA CEA code\footnote{http://www.grc.nasa.gov/WWW/CEAWeb/ceaHome.htm}\citep{gordon1996cea} at the prescribed planetary metallicity, derived by rescaling the solar elemental abundances \citep{asplund}.
For HD~209458b (0.7 M$_{Jup}$), we prescribe solar metallicity since it is close to one Jupiter mass.
For HAT-P-12b (0.2 M$_{Jup}$), we prescribe a planet metallicity of 22 $\times$ solar using the relation from \citet{kreidberg2014wasp43}.
In both cases, we reduce the oxygen volume mixing ratio to account for the rainout out of MgSiO$_3$ and MgSiO$_4$ which are below their condensation temperatures \citep{sing2016continuum}, but this only results in a $\sim$ 20\% change in the Oxygen abundance.

\begin{figure*}[!t]
\centering
\subfloat[HD 209458b]{
	\includegraphics[width=0.5\textwidth]{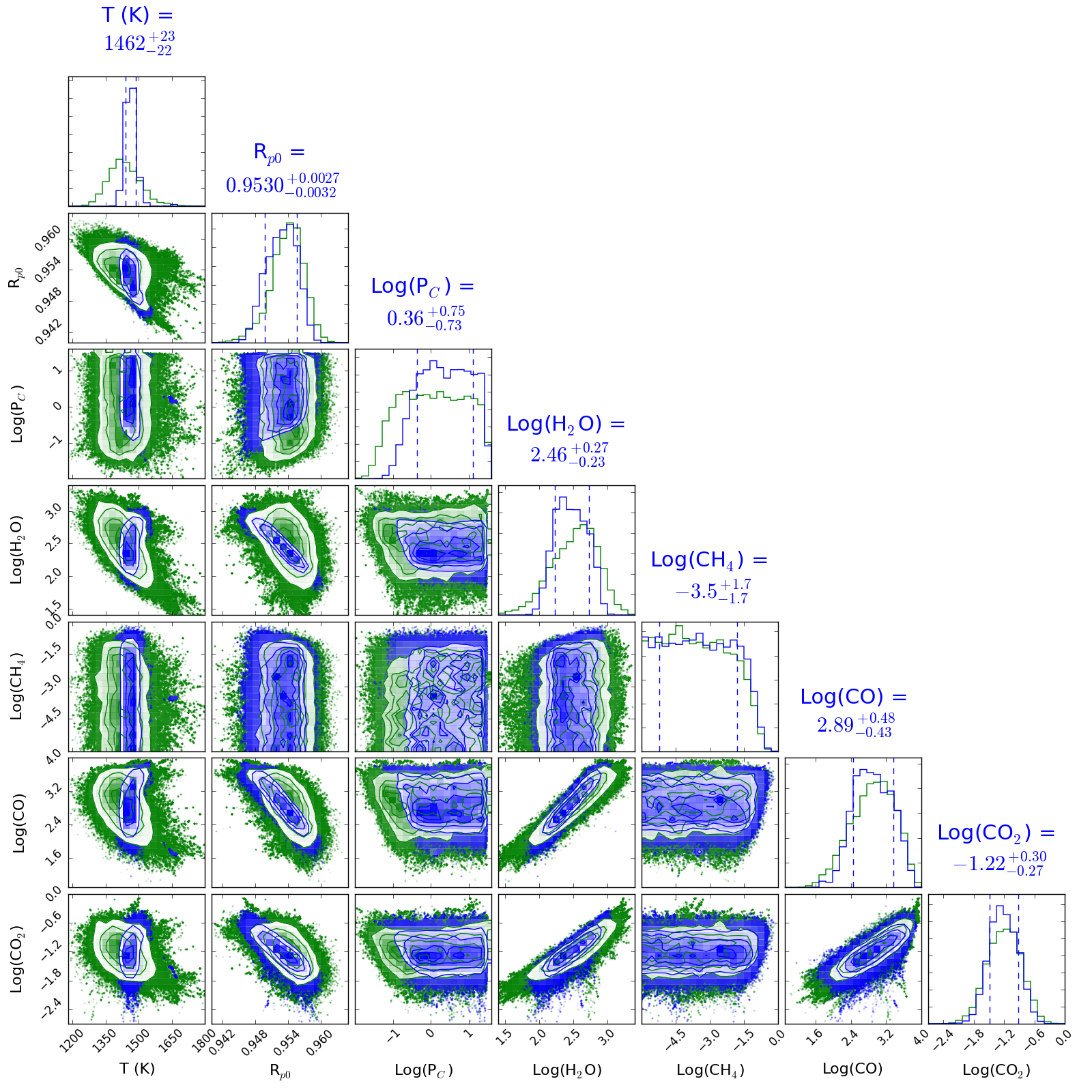}
	\label{fig:corner209}
	}
\subfloat[HAT-P-12b]{
	\includegraphics[width=0.5\textwidth]{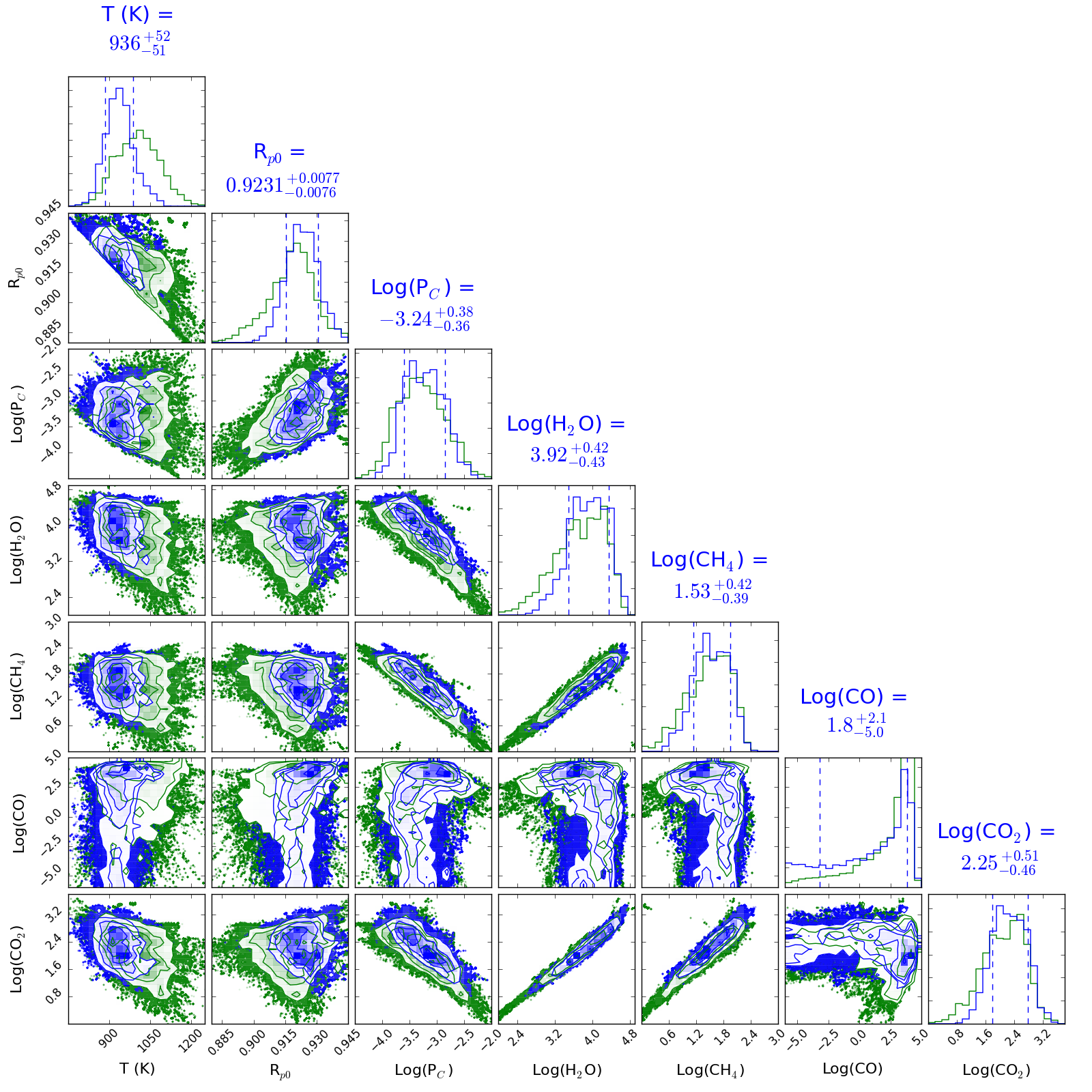}
	\label{fig:cornerHATp12}
	}
	\caption{Posterior distributions for MCMC runs of HD~209458b and HAT-P-12b with the DHS (blue) and without the DHS (green). The H$_2$O correlates strongly with CO$_2$ and CO, meaning that improved measurements of H$_2$O with the DHS actually help improve constraints on the Carbon bearing species, even though there aren't any significant CO or CO$_2$ features in the 1-2$\mu$m range. Abundances are given in $\log_{10}$~(ppm). Cloud pressure is in $\log_{10}$~(bar).}
	\label{fig:corner}
\end{figure*} 

\begin{figure}[!b]
\centering
\includegraphics[width=0.5 \textwidth]{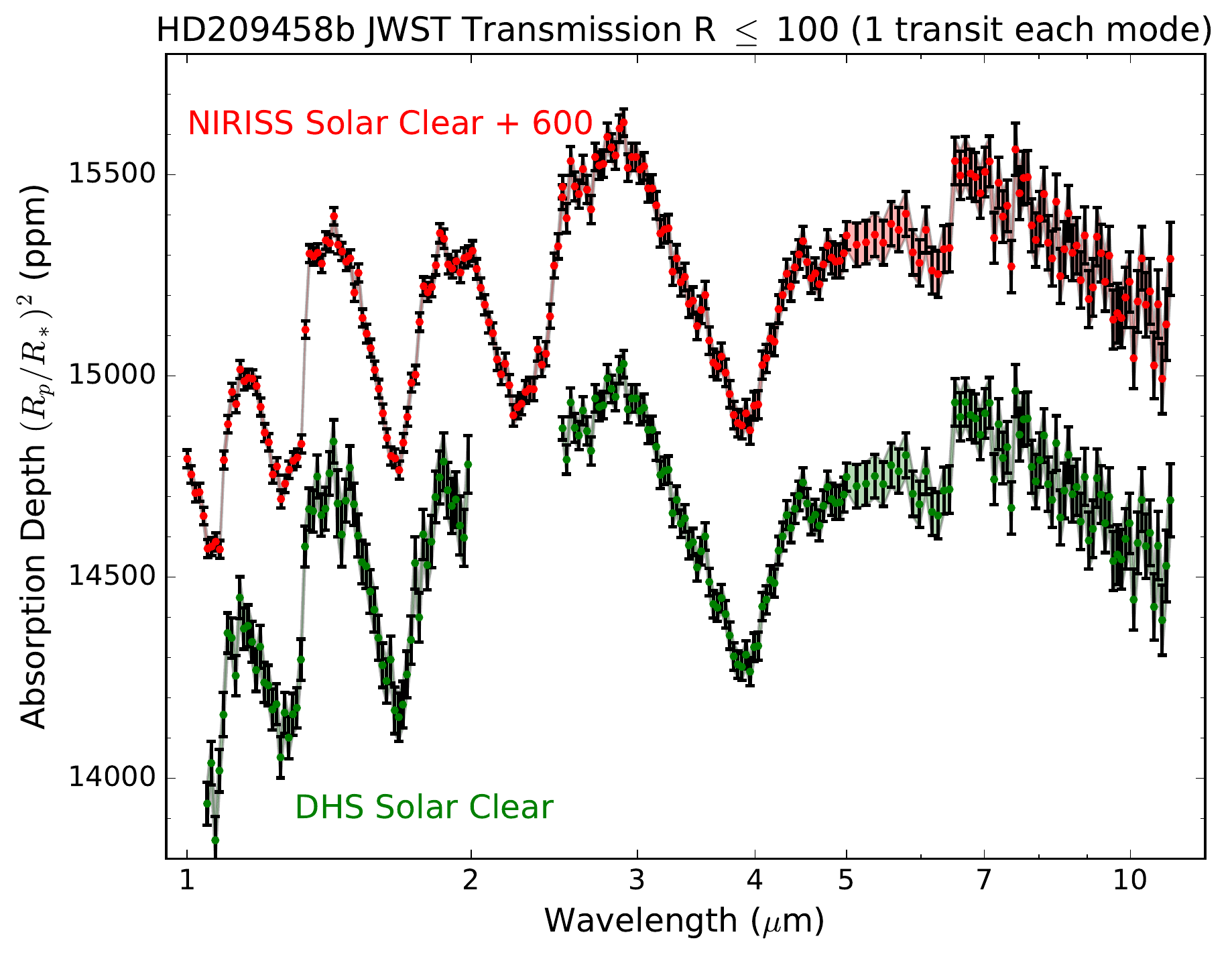}
\caption{Calculated SNR for one transit of the hot Jupiter HD~209458b for the NIRISS SOSS mode, NIRCam F322W2 and F444W filters and the MIRI LRS mode for a total of 4 different transits or $\sim$44 JWST hours.
Similar wavelength coverage can be achieved in just 3 different transits or $\sim$33 hours if the DHS mode is enabled (bottom spectrum).}\label{fig:DHSvsNIRISS209}
\end{figure}

We assume isothermal temperatures of 1440K and 960 K for HD~209458b and HAT-P-12b respectively. Molecular abundances are assumed constant with altitude at the 0.01 bar thermochemical values. While this ``0D" model is likely unrealistic, more sophisticated 1D models in general may be unrealistic given the uncertainties due to the impact of 3-dimensional effects \citep[e.g.][]{line2016nonUclouds,feng2016nonUniform}. We include H$_2$O, CO, CO$_2$, CH$_4$, NH$_3$, C$_2$H$_2$, HCN and collision-induced absorption (CIA) of hydrogen as the sources of opacity (with absorption cross-section derived from \citet{freedman2014opacities} and references therein), but only H$_2$O, CH$_4$, CO and CO$_2$ contribute significantly to the opacity, as seen in Figures \ref{fig:209spec} and \ref{fig:HATp12spec}.
For the HAT-P-12b model, we include an opaque gray cloud deck at 1 mbar to simulate the effect of clouds on reducing spectral features in a planet spectrum.
For HD~209458b, this cloud deck is placed at a pressure level of 1 bar which is too deep to affect the spectrum, as shown in Figure~\ref{fig:209spec}.

\subsubsection{Simulated Observations and Retrieval Methods}

For the simulated observations, we assume one transit is obtained in each of NIRCam's long-wavelength filters (F322W2 and F444W) to cover the wavelength range from 2.44$\mu$m to 4.98$\mu$m.
We assume that each of these was accompanied by a DHS observation for the 1.0$\mu$m to 2.0$\mu$m wavelength range, which benefits in SNR by $\sqrt{2}$ for the two separate observations.
This observing scenario is compared to a scenario that does not include the DHS to show the added scientific value of using the DHS spectra.
The readout mode in all cases simulated is RAPID readout (all reads saved up the ramp without skips, discussed in Section \ref{sec:readout}) with 10 reads up the ramp for HD~209458 and 25 reads up the ramp for HAT-P-12b.

The noise models for the hot Jupiter HD~209458b are shown in Figure~\ref{fig:DHSvsNIRISS209}, which is dominated by the residual systematic noise floor (set to 30 ppm) for the long-wavelength grisms ($>$2.44$\mu$m) but photon noise is more significant for the short-wavelength DHS observations.
The DHS has lower total throughput ($\sim$8\% including pupil coverage factor) and a smaller wavelength coverage than the NIRISS instrument in SOSS mode ($\sim$30\% throughput\footnote{\texttt{http://maestria.astro.umontreal.ca/niriss/simu1D/simu1D.php}}), so observations which require higher SNR and full wavelength coverage from \SOSSrangeto\ can take advantage of NIRISS (also shown in Figure~\ref{fig:DHSvsNIRISS209}) at the expense of more telescope time.

We use the \texttt{emcee} version of CHIMERA \citep{greene2016jwst_trans} to recover the individual atmospheric abundances of the simulated spectrum. The retrieval model includes 13 free parameters: 1) the temperature of the isothermal profile; 2) the altitude of the 10-bar pressure surface where the model begins; 3) a cloud pressure level below which all stellar flux is absorbed at all wavelengths; 4) through 11) the constant-with-altitude mixing ratios of H$_2$O, CH$_4$, CO, CO$_2$, NH$_3$, C$_2$H$_2$, HCN, N$_2$, respectively; 12) an amplitude of a Raleigh-like haze; and 13) a power law index for the Raleigh-type haze. We assume uniform-in-log priors from 10$^{-12}$ to 1 for all volume mixing ratios. The autocorrelation time for the retrievals is about ~100 steps and we run all retrievals to 20,000 steps, only using the last 5,000 (times 160 walkers) to allow for burn-in.


\subsection{Parameter Estimation Results}\label{sec:retrievals}

The posterior distributions for retrieved parameters for the most important abundances are shown in Figure~\ref{fig:corner}.
We show the posterior distributions for two cases:
\begin{enumerate}
\item \label{it:grAlone} An observation with the long wavelength grisms alone (2.4-5.0$\mu$m), shown in green.
\item \label{it:DHSgr} An observation with the long wavelength grisms (2.4-5.0$\mu$m) with simultaneous DHS spectra (1.0-2.0$\mu$m), shown in blue.
\end{enumerate}
For the spectral data they have in common (2.4$\mu$m to 5.0$\mu$m), we use the same noise instance to make direct comparison easier.

The posterior distributions for HD~209458b in Figure~\ref{fig:corner209} show improved constraints on the temperature, cloud pressure level and the abundances of H$_2$O, CO and CO$_2$ when the DHS is used simultaneously with the long wavelength grisms (Case~\ref{it:DHSgr}) as compared to when the long wavelength grisms are used alone (Case~\ref{it:grAlone}).
These improvements come about, in part, because the 1.0$\mu$m to 2.0$\mu$m wavelength region probes to deeper pressure layers ($\sim$ 400 mbar) to exclude clouds and better constrain the scale height.


The warm Saturn HAT-P-12b model shows improvements to the temperature constraints, but the cloud pressure measurements have similar uncertainties with and without the DHS.
This is because the 1mbar cloud layer begins to affect the opacity near 4.0$\mu$m, which is measured by the long wavelength grisms.
As with HD~209458b, the DHS improves constraints on the H$_2$O and CO$_2$ abundances, but since the planet is at a cooler temperatures, chemical equilibrium favors CH$_4$ over CO, so the constraints are much better on CH$_4$.


One of the surprising aspects to the retrievals is that the DHS improves both the constraints on the CO and CO$_2$ abundances as well as the H$_2$O abundances, seen in Figure~\ref{fig:corner209} and Figure~\ref{fig:cornerHATp12}. 
A quick glance at the contributions of individual molecules to the spectrum shown in Figures \ref{fig:209spec} and \ref{fig:HATp12spec} would indicate that the DHS is mainly important for water vapor.
However, the retrieval for the spectrum with only NIRCam's long-wavelength grisms has a strong correlation between the retrieved H$_2$O abundances with CO$_2$ and CO, shown in Figure~\ref{fig:corner209}.
Improving the water vapor constraint with the DHS, reduces the correlation and therefore improves the abundance constraints for all three.

\section{Conclusions}

The James Webb Space Telescope will provide unprecedented measurements of transiting exoplanets, especially for carbon-bearing and nitrogen-bearing gases in their atmospheres.
The time needed to observe in JWST's full range of wavelengths from 0.6$\mu$m to 28$\mu$m, however, can add up quickly since many transits are needed with different instrument modes to cover all wavelengths.
We introduced a new science mode using the Dispersed Hartmann Sensors on the NIRCam instrument which enable simultaneous measurements of spectra at R $\sim$ \DHSres\ from 1.0 to 2.0$\mu$m while using the NIRCam long-wavelength grisms covering either 2.4$\mu$m to 4.0$\mu$m or 3.9$\mu$m to 5.0$\mu$m.
The fact that the two sets of grisms operate simultaneously allows more science to be done in the same amount of time.
The readout mode and subarray size must be chosen carefully on the DHS so as to not saturate the long-wavelength grisms.

The DHS mode enables more science because it has the brightest saturation limits of any JWST spectroscopic mode and can increase the wavelength coverage of a single observation.
The brightest known systems and similarly bright future TESS targets will be accessible with NIRCam's DHS mode.
We present example usage of the DHS mode on two systems and show that the temperature, cloud pressure level and water vapor constraints are improved with the DHS.
Furthermore, the improved water vapor constraints also feed back to higher precision CO and CO$_2$ abundances because of the correlation of these derived parameters.
We encourage the implementation and approval of the DHS mode to enable this additional science in less time for this limited life observatory.

\subsection{Specific Changes Needed for Implementation}
We summarize the specific changes to the existing operations needed for the DHS here:
\begin{enumerate}
\item Allow the NIRCam pupil wheel to be used for science at the DHS element position with spectra parallel with detector rows
\item Allow subarray locations to be different for the short wavelength arrays and the long wavelength arrays in order to capture the spectrum for element number 7 for 2048~$\times$~64 and 2048~$\times$~128 subarrays and both element numbers 7 and 8 for the 2048~$\times$~256 subarray
\item Choose field points from NIRCam's long wavelength grism to allow simultaneous DHS spectroscopy with all 10 DHS elements when reading out in full frame (2048~$\times$~2408) mode
\item Change the OSS software and timing so that large subarrays may used with the DHS simultaneously with small 2048$\times$64 subarrays with the long wavelength grisms
\end{enumerate}

\vspace{0.1in}
\section*{Acknowledgements}

CHIMERA retrieval makes use of \texttt{emcee} \citep{foreman-mackey2013emcee} and the covariance plot was made with \texttt{corner.py} \citep{foremanCorner}.
Funding for the NIRCam team is provided by NASA Goddard Spaceflight Center.
For use of the El Gato computing system, this material is based upon work supported by the National Science Foundation under Grant No. 1228509.
Thanks to Michael Bruck at the High Performance Computing center at the University of Arizona for assisting with setting up the CHIMERA runs to utilize the HPC hardware.
This research has made use of the Exoplanet Orbit Database and the Exoplanet Data Explorer at exoplanets.org.

\appendix




\bibliographystyle{apj}
\bibliography{dhs_biblio}

\end{document}